\documentstyle[11pt]{article}

\begin{document}
\oddsidemargin .3in \evensidemargin 0 true pt \topmargin -.4in

\def\ra{{\rightarrow}}
\def\a{{\alpha}}
\def\b{{\beta}}
\def\l{{\lambda}}
\def\eps{{\epsilon}}
\def\T{{\Theta}}
\def\t{{\theta}}
\def\co{{\cal O}}
\def\car{{\cal R}}
\def\caf{{\cal F}}
\def\cs{{\Theta_S}}
\def\pr{{\partial}}
\def\tri{{\triangle}}
\def\na{{\nabla }}
\def\S{{\Sigma}}
\def\s{{\sigma}}
\def\sp{\vspace{.15in}}
\def\hs{\hspace{.25in}}

\newcommand{\be}{\begin{equation}} \newcommand{\ee}{\end{equation}}
\newcommand{\bea}{\begin{eqnarray}}\newcommand{\eea}
{\end{eqnarray}}


\begin{titlepage}
\topmargin= -.2in \textheight 9.5in

\begin{flushright}
\end{flushright}
\baselineskip= 16 truept

\vspace{.3in}

\centerline{\LARGE {\bf Noncommutative $D_3$-brane, Black Holes}}

\centerline{\LARGE {\bf and Attractor Mechanism}}

\vspace{.4in}

\centerline{{{\Large Supriya Kar}${}\ ^{a,b,}$\footnote{skkar@physics.du.ac.in, skar@ictp.trieste.it }}
{{\Large and Sumit Majumdar}${}\ ^{b,}$\footnote{sumit@physics.du.ac.in}}}

\vspace{.2in}

\centerline{$^a${\large The Abdus Salam International Centre for Theoretical Physics}}
\centerline{Strada Costiera 11, Trieste, Italy}

\vspace{.2in}

\centerline{
$^b${\large Department of Physics \& Astrophysics,
University of Delhi}}
\centerline{\large Delhi 110 007, India }

\vspace{.2in}

\thispagestyle{empty}

\baselineskip= 17 truept

\begin{center}
({\tt {\today}})

\vspace{.3in}
\noindent {\bf Abstract}

\end{center}
\vspace{.1in}

\noindent We revisit the $4D$ generalized black hole geometries,
obtained by us \cite{km-2}, with a renewed interest, to unfold
some aspects of effective gravity in a noncommutative $D_3$-brane
formalism. In particular, we argue for the existence of extra
dimensions in the gravity decoupling limit in the theory. We show
that the theory is rather described by an ordinary geometry and is
governed by an effective string theory in $5D$. The extremal black
hole geometry $AdS_5$ obtained in effective string theory is shown
to be in precise agreement with the gravity dual proposed for
$D_3$-brane in a constant magnetic field. Kaluza-Klein
compactification is performed to obtain the corresponding charged
black hole geometries in $4D$. Interestingly, they are shown to be
governed by the extremal black hole geometries known in string
theory. The attractor mechanism is exploited in effective string
theory underlying a noncommutative $D_3$-brane and the macroscopic
entropy of a charged black hole is computed. We show that the
generalized black hole geometries in a noncommutative $D_3$-brane
theory are precisely identical to the extremal black holes known
in  $4D$ effective string theory.

\vspace{.3in}




\end{titlepage}

\baselineskip= 17 truept

\section{Introduction}

Black holes are macroscopic objects with strong curvatures
in space-time and they possess a non-zero temperature. Naturally, the
thermodynamic properties of the black holes are characterized by its
macroscopic entropy. In fact, the nature of thermodynamic entropy of
a black hole is very similar to that of the Bekenstein-Hawking. It
is known to be governed by one quarter of its area of the
horizon in Planck units \cite{bekenstein,hawking}. Thus, on the one
hand, the computation of thermodynamic entropy of a black hole
involves the counting of microstates, which is based on statistical analysis.
On the other hand, the Bekenstein-Hawking entropy of a black hole appears to
be governed by some macroscopic interpretations.

\sp
\par
In the recent past, the issue of microscopic analysis relevant for
the computation of entropy of a black hole was revisited in string
theory \cite{strominger}-\cite{ooguri}. Interestingly, the
microscopic entropy was computed for a certain magnetically charged
black holes in its near horizon geometry \cite{ferrara} following an
attractor mechanism in the theory. There, the variation of the
moduli fields in string theory is governed by the damped
geodesic equation on the moduli space. The damping is essentially
caused by the presence of the electromagnteic field in the theory.
The geodesic equation possesses an attractive fixed point at its
event horizon. Since the area of the event horizon is determined
precisely by the electromagnetic charges, the variation of the
asymtotic moduli fields does not affect the near horizon geometry of
the macroscopic black hole. In other words, the number of internal
black hole states remain unchanged under the influence of an
adiabatic change in its macroscopic environment. It leads to the
statistical interpretation  of the Bekenstein-Hawking entropy.

\sp
\par
On the other hand, there are renewed interests to investigate some
of the related issues in the quantum gravity with the developments
of nonlinear electrodynamics on a $D$-brane
\cite{seiberg-witten}-\cite{horo-pol},\cite{km-2}. The stringy formulations have
indeed motivated the construction of some of the realistic
brane-world models, which are known to describe various effective
theories of gravity. Interestingly, the construction of $D$-brane
solutions leading to solitons, shock waves and black holes have been
obtained in the folklore of string theory
\cite{tseytlin,gibbons-hashi,gibbons-herdeiro,gibbons-ishibashi,nasseri,
kaloper,km-1,nasseri2,km-2}.
Thus, in a brane-world scenario, one needs a better understanding of
the effective nature of gravity derived from the nonlinearity in the
elctromagnetic (EM-) field.

\sp
\par
In the context, the computation of black hole microstates has been
formalized by Wald \cite{wald}. Interestingly, the
Bekenstein-Hawking entropy of an extremal black hole has been computed
in presence of various different higher derivative terms in string
theory \cite{dhabolkar}-\cite{alisha}. For instance, in an arbitrary
$d$-dimensions, the near horizon geometry of these black holes are
governed by $AdS_2\times S^{d-2}$. Then, the black hole entropy is
shown to be defined as a function of electric and magnetic charges,
respectively, associated with the one-form and the ($d-3$)-form
gauge fields in theory \cite{sen}.

\sp
\par
In this paper, our primary motivation is to investigate some aspects of an effective theory of gravity, formulated, in
a noncommutative frame-work \cite{jevicki}-\cite{alvarez}. Very recently,
one such attempt was made by us in a formulation based on a noncommutative $D_3$-brane \cite{km-2}. Generalized Reissner-Nordstrom (RN-) and Schwarzschild black holes in $4D$
are obtained in the effective frame-work describing a curved $D_3$-brane. Most importantly,
the generalized RN-black hole geometry obtained is new in the effective string theory. This
is similar to the case for Einstein-Maxwell black hole with a generalized mass and a charge.
A priori, the generalized RN-black holes are different than the charged black holes obtained
in string theory \cite{garfinkle}. This is due to the fact that the dilaton couples to the
gauge field strength in string theory. As a result, every solution in presence of a gauge field
in string theory must possess a nonconstant dilaton. On the contrary, the generalized RN-black holes are independent of the value of the dilaton in effective string theory. It provides
hint towards an underlying attractor mechanism for the black holes in a noncommutative frame-work.
In particular, we see that the generalized black hole geometries in a noncommutative $D_3$-brane theory are precisely identical to the ones obtained in effective string theory with an ordinary geometry \cite{garfinkle}.

\sp
\par
In the context, we compute the entropy function of a generalized
black hole. It is argued that the attractor mechanism established in string theory
\cite{ferrara,strominger,ferrara2} may be extended to an effective
string formulation based on a noncommutative $D_3$-brane. As a
result, the $4$-dimensional generalized black holes \cite{km-2} are
further investigated to explore the possibility of extra dimensions,
if any, in the Planckian regime. Working out the constraints arising
out of the noncommutativity, it is shown that the effective string
theory governs an ordinary geometry in $5$-dimensions. The
additional small dimension in the theory is essentially due to the
curved nature of $D_3$-brane and is transverse ($\perp$-) to its
flat world volume. Intuitively, the new $\perp$-dimension seemingly
traverses into the bulk of the string from its boundary. Alternately,
the bulk description is argued to govern an effective string in
$5$-dimensions. Various black hole solutions characterized by the
effective mass $M_{\rm eff}$ and charge $Q_{\rm eff}$ are obtained
in the frame-work, which are based on an underlying space-time
noncommutativity on a $D_3$-brane at its gravity decoupling limit.
Interestingly, in the limit, a black hole is governed by its near horizon
geometry and is precisely described by the one obtained \cite{li-wu} using $AdS_5$
/ noncommutative Yang-Mills correspondence.
Furthermore, we perform Kaluza-Klein compactification of effective
string theory and obtain $4D$ extremal black hole goemetry in string and Einstein
frames. Though, the extremal black holes are
obtained in an effective theory of gravity, their geometrical fate
is purely dependent on the nonlinear EM-field in the frame-work.
The extremal black holes resemble to that obtained in
string theory \cite{garfinkle}. At the first sight, we would like to keep a note
that our analysis is in agreement with the
$AdS/CFT$ conjecture established in string theory \cite{hashi-itzhaki,li-wu,ho-li,maldacena2}.
Most importantly, the $4D$ black holes obtained in a Kaluza-Klein compactified theory in the bulk are in precise correspondence with the ones obtained on a curved $D_3$-brane in its gravity decoupling limit. In otherwords, the black hole geometries in an effective string theory are identical to that of the generalized geometries obtained in a noncommutative $D_3$-brane theory.

\par
\sp
The plan of this paper is as follows. In section 2.1, we outline
the relevant results from our recent work \cite{km-2}. The possibility
of extra dimensions in a formalism based on noncommutative $D_3$-brane in its
gravity decoupling limit is argued in section 2.2. The required effective
string description in $5D$ is obtained in
section 2.3. Subsequently, the black hole geometries are constructed in
$5D$ and is shown to be in precise agreement with the result
obtained in a different context using $AdS_5$ / noncommutative gauge
theory correspondence. The relvant $4D$ extremal black
hole geometries in Einstein and string frames are obtained, respectively, in sections
3.1 and 3.2 by using the Kaluza-Klein compactifcation in the frame-work.
In section 3.3, the attractor mechanism is analyzed in the noncommutative $D_3$-brane
formalism to compute the black hole entropy function. Finally, we conclude with some remarks
in section 4.

\section{Noncommutative $D_3$-brane and the notion of effective string}

\subsection{Preliminaries}

Consider a $D_3$-brane, in presence of a constant two form $b$
induced on its world-volume. The uniform EM-field on the $D_3$-brane
is governed by a nonlinear electrodynamics. Our starting point is in
type IIB string theory. In principle, the gravity and the gauge
dynamics in the frame-work may be approximated by coupling the
$D_3$-brane dynamics to a generalization of Einstein's action,
$i.e.$ in presence of higher derivatives terms. Since the present
work is primarily confined to the gravity decoupling limit in the
theory, the higher derivative terms in the gravity sector shall not
contribute significantly in the frame-work.

\par
\sp
In a static gauge for the space-time, the bulk metric may be
viewed on the world-volume and the complete action becomes
\cite{km-2}
\be
S= \int d^4y\ {\sqrt{g}}\ \left (\ {1\over{16\pi
G_N}}\ R \ - {1\over4} g^{\mu\nu}g^{\lambda\rho}\ {\cal
F}_{\mu\lambda} {\cal F}_{\nu\rho} \ + \ {\cal O}({\cal F}^4)\ +\
\dots\ \right )\ ,\label{pre-1}
\ee
where the $U(1)$ gauge field
${\bar{\cal F}}_{\mu\nu}= (b + 2\pi\alpha'F)_{\mu\nu}$. In
absence of higher derivative terms in gauge field, the frame-work
resembles to the Einstein's theory coupled to the Maxwell's.
However in presence of higher derivatives, the action
(\ref{pre-1}) can be given by
\be
S=\ \int d^4y\ {\sqrt{g}}\ \left
(\ {1\over{16\pi G_N}}\ R \ - {1\over4} \Big [\ {\cal F}^2 -\
{1\over2} {\cal F}{\cal F}_{+}\ F_{-}^2\ K^2\big ( {\cal F}\big )\
\Big ]\ \right ) \ ,\label{pre-2}
\ee
where ${\cal F}_{\pm}= ({\cal F}\pm {}^{\star}{\cal F})$ and $K\big ({\cal F}\big )$
contains all the higher order terms in field strength. The Hodge dual of ${\cal F}$
is denoted as ${}^{\star}{\cal F}$. The Minkowski's inequality
can be seen to  yield the (anti-) self-duality condition $|{\bf E}|=
|{\bf B}|$ in the theory. Since all the higher order terms in gauge
fields vanish, the frame-work leads to an exact stringy description.
Then, the relevant action on a curved $D_3$-brane becomes
\be
S= \int d^4y\ {\sqrt{g}}\ \Big (\ {1\over{16\pi G_N}}\ R  \ - {1\over4}
g^{\mu\lambda} g^{\nu\rho}\ {\cal F}_{\mu\nu}{\cal F}_{\lambda\rho}\
\Big )\ .\label{pre-3}
\ee
Interestingly, the Einstein's equation
is governed by the vacuum equations $i.e.\ T_{\mu\nu} = 0$, which is
due to the self-dual nonlinear gauge field in the theory.

\par
\sp
Now let us consider an equivalent noncommutative gauge dynamics
on the $D_3$-brane \cite{seiberg-witten}. The ordinary product
is replaced by the Moyal $\star$-product on the world-volume, which introduces nonlocal
terms in the gauge theory due to the infinite number of derivatives there.
However, it does not affect the bulk dynamics
$i.e.$ the gravity sector. Then the action (\ref{pre-3}), for a
curved $D_3$-brane, can alternately be given by
\be
S\ =\int d^4y\
{\sqrt{G}} \left ( {1\over{16\pi G_N}} R\ -  {1\over4}
G^{\mu\lambda} G^{\nu\rho}\ {\hat F}_{\mu\nu} \star {\hat
F}_{\lambda\rho}\ \right )\ .\label{pre-4}
\ee
where $G_{\mu\nu}$ denotes the effective metric
It can be checked that the $G_{\mu\nu}$ can be generalized to include higher order terms in
the two-form potential $b$. On can re-express the effective metric as
\be
G_{\mu\nu} = g_{\mu\nu }\ - \left(bg^{-1}b\right)_{\mu\nu}\ + \ \left ([bg^{-1}b]
[bg^{-1}b]\right )_{\mu\nu}\ + \ \dots\ \ .\label{5des-300}
\ee
With a gauge choice $G_{i\a}=0$ for
$(\a,\b) = (y^0,y^1)$ and $(i,j) =(y^2,y^3)$, the action can be
simplified using a noncommutative scaling \cite{km-2}.{\footnote{The scaling may also be seen as
that of Planckian energy limit as discussed in refs.\cite{ver-2,kar-maharana}.}}
The action takes a form
\be
S\ = \int d^2y^{(\alpha)} d^2y^{(i)}
{\sqrt{\bar h}}\ {\sqrt{h}}\; \Big [ {1\over{16\pi}}R_h +
{1\over{64\pi}}\ h^{ij}\ \partial_i {\bar h}_{\alpha\beta}
\partial_j{\bar h}_{\gamma\delta}
\epsilon^{\alpha\gamma}\epsilon^{\beta\delta} - {1\over2} {\bar
h}^{\alpha\beta} h^{ij}\ {\hat F}_{\alpha i}\star {\hat F}_{\beta
j}\ \Big ) \Big ] \ ,\label{pre-5}
\ee
where $h_{ij}$ and ${\bar h}_{\a\b}$ denote the components of $G_{\mu\nu}$, respectively, in
the $\perp$- and longitudinal ($L$-) spaces. The action is derived by using the vacuum field
configurations, $i.e.\ \partial_{\alpha}h_{ij}= 0$, $R_{\bar h} = 0$ and
${\hat F}_{\alpha\beta}=0$.

\par
\sp
Unlike to the nonlinear EM-theory
(\ref{pre-2}), the energy-momentum tensor turns out to be
significant in the noncommutative frame-work (\ref{pre-5}).
The generalized black hole geometry in the theory can be
derived. The computational details are beyond the scope of
this paper and may be checked from our recent work \cite{km-2}.
Finally, the generalized black hole geometry is given by
\bea
ds^2&=& -\left (1-{{ 2M_{\rm
eff}}\over{r}} - {{Q^2_{\rm eff}}\over{r^4}} + {{2M_{\rm eff}\
Q^2_{\rm eff}}\over{r^5}}\right ) dt^2 + \left (1-{{2M_{\rm
eff}}\over{r}} - {{Q^2_{\rm eff}}\over{r^4}} + {{2M_{\rm eff}\
Q^2_{\rm eff}}\over{r^5}} \right )^{-1} dr^2\nonumber\\
&&\quad\qquad\qquad\qquad\qquad\qquad + \left ( \ 1 - {{Q^2_{\rm
eff}}\over{r^4}} \right ) r^2 d\theta^2 \ +\ \left ( \ 1 -
{{Q^2_{\rm eff}}\over{r^4}} \right )^{-1} r^2 \sin^2 \theta \ d\phi^2
\ ,\label{pre-6}
\eea
where mass and charge are generalized due to
the noncommutative $\Theta$-terms. Explicitly, they are given by
\bea
&&\quad M_{\rm eff}= (G_NM)\left [ 1 - {{\Theta}\over{2r^2}} +\
{\cal O}(\Theta^2)\ +\ \dots\ \right ]
\qquad\qquad\qquad\qquad\nonumber\\
{\rm and}&&\quad {Q^2_{\rm eff}}= (G_N Q^2) \left [1 -
{{\Theta}\over{r^2}} +\ {\cal O}(\Theta^2)\ +\ \dots\ \right ]\ .\label{pre-7}
\eea

\subsection{Extra dimensions}

The generalized black hole geometry \ref{pre-6}) in its gravity decoupling limit,
$i.e.\ M\rightarrow 0$, gives rise to a Schwarzschild-like geometry.
It is given by
\be
ds^2= \left (1 -{{Q^2_{\rm eff}}\over{r^4}}\right ) \Big [- dt^2\
+ r^2 d\theta^2\Big ] \ +\ \left (1- {{Q^2_{\rm eff}}\over{r^4}}
\right )^{-1} \Big [ dr^2 + r^2 \sin^2 \theta\ d\phi^2\Big ]\ .\label{extra-1}
\ee
In the limit, the black hole is characterized by its charge, which
can also be interpreted as its light mass in the frame-work. In fact,
the correction term $Q^2_{\rm eff}/r^4$ is essentially due to the non-zero energy-momentum tensor and hence is associated with the mass
of the black hole. Since an $n$-dimensional Schwarzschild black hole mass is known to be associated with $r^{3-n}$ term in its metric
component $G^{tt}$, a priori, the black hole (\ref{extra-1}) can be seen to be governed by a $7$-dimensional space-time. However, these dimensions are not all independent due to the noncommutative constraints in the
frame-work. The number of constraints on the space-time degrees of freedom can be computed from the EM-field configurations in the theory. For instance, two different length scales in the
frame-work can be argued with two non-zero components of the $E$-field and one
non-zero component of $B$-field or vice-versa. The self dual EM-field reduces the number of
noncommutative constraints to two. As a result, the resulting $7D$ space-time effectively governs a $5D$ space-time with ordinary geometry. The othogonality in the space-time enforces that the extra dimension is transverse to the $D_3$-brane world-volume. A priori, the world-volume coordinates together with three of the extra dimensions describe an effective string in $5D$ with an ordinary geometry. In otherwords,
the noncommutativity is used to separate two large $\perp$-dimensions ($y^2,y^3$) scaled apart from the longitudinal ones ($y^0,y^1$). The presence of three small dimensions, in the regime, make the total dimension of space-time to five.

\subsection{Effective string in $5D$ and charged black holes}

We begin this section with an effective string dynamics in
$5$-dimensions. Since the notion of a curved $D_3$-brane is contained in a
type IIB string theory, we consider the string compactified on $K^3\times S_1$.
For instance, the generic form of
the effective string action may be obtained from ref.\cite{kmp} and
is given by
\bea
S&=&\int d^5 x \sqrt{-\cal{G}} e^{-\Phi} \left [ {\cal{R}} +
(\partial\Phi)^2-{1\over4}{{\cal{F}}^{(\tilde{m})}}{{C_{\tilde{m}\tilde{n}}}}{{\cal{F}}^{(\tilde{n})}}
- {1\over{12}} {\cal{H}}^2\right ] ,\label{5des-1}
\eea
where
${\cal{G}}=\det {\cal G}$ and $\Phi$ denotes the dilaton. ${\cal
F}^{(\tilde{m})}$ is a two form and ${\cal H}$ is a three form gauge field strength.
$C_{\tilde{m}\tilde{n}}$ is a square matrix and it signifies the appropriate moduli field couplings in the
theory. The irrelevant moduli term, $i.e.\ M_{pq}\partial_{\a}\phi^p\partial^{\a}\phi^q$,
is dropped from the action(\ref{5des-1}). In the Einstein frame, $i.e.\ {\cal
G}_{\a\b}=e^{2\Phi/3}{\cal G}^E_{\a\b}$, the action may be
re-expressed as \bea S&=&\int d^5 x \sqrt{-{\cal{G}}^{E}}\left [
{\cal{R}}-{1\over3}(\partial\Phi)^2-
{1\over4}e^{-2\Phi/3}{\cal{F}}^{(\tilde{m})}C_{\tilde{m}\tilde{n}}
{{\cal{F}}^{(\tilde{n})}}-{1\over{12}}e^{-4\Phi/3}{\cal{H}}^2 \right ] .\label{5des-2}
\eea
The action can be re-expressed using the duality
\be
e^{{-\Phi}/3}H^{\mu\nu\l}={{\epsilon^{\mu\nu\l\s\rho}}
\over{2!\sqrt{-{\cal{G}}^E}}} \tilde{F}_{\s\rho} \ .\label{5des-3}
\ee
It is given by
\bea
S&=&\int d^5 x \sqrt{-{\cal{G}}^{E}}\left [
{\cal{R}}-{1\over3}(\partial\Phi)^2-
{1\over4} {\cal{F}}^{(m)}\Lambda_{mn}[\Phi] {{\cal{F}}^{(n)}}
\right ]  \ ,\label{5des-4}
\eea
where
\bea \Lambda_{mn}[\Phi]= e^{-{2\Phi}/3}\left( \matrix{ {C_{\tilde{m}\tilde{n}}} & 0 \cr 0
& 1} \right)\quad {\rm and} \qquad {\cal{F}}^{(m)}=\left(
\matrix{{\cal{F}}^{(\tilde{m})} \cr \tilde{F} }
\right)\ .\label{5des-5}
\eea
Though the EM-field,
$\textbf{E}=(0,E_2,E_3)$ and $\textbf{B}=(0,B_2,B_3)$, are constants
on a $D_3$-brane world-volume, their generalization to the
$5$-dimensional string bulk leads to nonconstant $E$- and $B$-
fields. The E-field can be generalized appropriately, by using the effective metric (\ref{5des-300}),
to yield a redefined electric field $\tilde{\mathbf E}$ on the world-volume.
It is given by
\be
{\tilde{\mathbf E}}^2= {{{\mathbf E}^2}\over{1+{{\mathbf E}^2}}}\ .\label{5des-301}
\ee
Now the ansatz for a nontrivial
$5$-dimensionsal string metric may be constructed and is given by
\be
ds^2 =\ dr^2\ +\ G_{\mu\nu}(r)\ dx^{\mu}dx^{\nu}\ .\label{5des-6}
\ee
Using a Schwarzschild black hole geometry,
$i.e.$ a vaccum solution for the genuine metric $g_{\mu\nu}$
(\ref{pre-6}), the $5$-dimensional solution to the effective string
theory (\ref{5des-4}) is worked out to yield
\bea
ds^2&=&dr^2+\Bigg
[ -\left (1 - {{2M_{\rm eff}}\over{r}}\right )(1-E^2)\ dt^2 + \left
( 1-{{2M_{\rm eff}}\over{r}}\right)^{-1}(1-E^2)^{-1} d\rho^2
\nonumber\\
&&\qquad\qquad\qquad\quad\qquad\qquad +\left ( 1+B^2 \right
)\rho^2d\theta^2 + \left ( 1+B^2 \right )^{-1} \rho^2 \sin^2\theta\
d\phi^2 \Bigg ] \ ,\label{5des-7}
\eea
where $r$ denotes the
$\perp$-coordinate to the $D_3$-brane world-volume ($t,\rho,\theta,\phi$). Incorporating
the self-duality of the EM-field, we re-express the metric in terms of
the isotropic spherical coordinates
for our purpose. Then, the $5D$ string solution in
$(t,u,\rho,\theta,\phi)$-coordinates, for $u=1/r$, becomes
\bea
ds^2&=&{{du^2}\over{u^2}} \ + \ u^2  \Bigg (\ \left [-\left (
1-{{M_{\rm eff}}\over{2r}}\right )^2 \left ( 1+ {{M_{\rm eff}}\over{2r}}\right )^{-2}\ +\
{\tilde E}^2\left ( 1+{{M_{\rm eff}}\over{2r}}\right )^{-4}\right ] dt^2\nonumber\\
&& \qquad\qquad\qquad +\ \left [\left ( 1+{{M_{\rm eff}}\over{2r}}
\right )^4 -\ {\tilde E}^2
\left ( 1+{{M_{\rm eff}}\over{2r}}\right )^2 \left ( 1-{{M_{\rm eff}}\over{2r}}\right )^{-2}\right ] d\rho^2\nonumber\\
&& \qquad\qquad\qquad\qquad\qquad +\ \left [\left ( 1+ {{M_{\rm
eff}}\over{2r}} \right )^4-\ {\tilde E}^2 \left ( 1+ {{M_{\rm eff}}\over{2r}}
\right )^{-4}\right ]\rho^2\ d\Omega^2 \Bigg ) \ ,\label{5des-71}
\eea
where ${\tilde E}$ denotes the $E$-field in $5$-dimensions. In the
gravity decoupling limit, $i.e.\ M\rightarrow 0$, the geometry
simplifies drastically to yield
\be
ds^2={{du^2}\over{u^2}}\ +\ u^2\left ( 1- {\tilde E}^2\right ) \left [ -dt^2+d\rho^2+\rho^2d \Omega ^2\right ]\ .\label{5des-8}
\ee
Using eq.(\ref{5des-301}), the
$E$-field in the theory can be given by
\be
{\tilde E} = \left [{{Q_{\rm eff}^2u^4}\over{1+Q_{\rm eff}^2 u^4}}\right ]^{1/2}\ .\label{5des-9}
\ee
It is straight-forward to check that for large $u$, the constant value of the EM-field is recovered, which corresponds to that on the $D_3$-brane. The IR-limit, $i.e.\ u\rightarrow 0$, corresponds to the string bulk and in the UV-limit,
$i.e.\ u\rightarrow \infty$, defines the string boundary.
Interestingly, the gravity decoupled solution
(\ref{5des-8})-(\ref{5des-9}) obtained following a noncommutative
$D_3$-brane formulation describes an extremal black hole there. It
is precisely correspond to the gravity dual proposed for
$D_3$-brane in presence of a constant magnetic field \cite{li-wu}.
In other words, it provides evidence for the holographic
correspondance between the bulk of the string and the noncommutative
$U(1)$ theory on its boundary.

\sp
\noindent
Re-writing the extremal black hole geometry
(\ref{5des-8}) in $(t,r,\rho,\theta,\phi)$-coordinate, one gets
\be
ds^2={{dr^2}\over{r^2}}+{{r^2}\over{Q_{\rm eff}^2+r^4}}\left [
-dt^2+d\rho^2+\rho^2d\Omega^2\right ]\ . \label{5des-99}
\ee
Interestingly, the geometry remains unchanged under a change
$r\rightarrow 1/r$. In $(t,u,\rho,\theta,\phi)$-coordinate system,
the extremal black hole geometry governs a weekly coupled gravity in
the string bulk and a strongly coupled noncommutative $U(1)$ gauge
theory at its boundary. The situation reverses in the
$(t,r,\rho,\theta,\phi)$-coordinate system, $i.e.$ a weekly coupled
gauge theory at the string boundary and a strongly coupled gravity
in its bulk. It further reconfirms the strong-week coupling string
duality between the (noncommutative) gauge and (ordinary) gravity theories.

\sp
\noindent
Alternatively, the extremal limit can be incorporated
by taking $M_{\rm eff}\rightarrow Q^{2}_{\rm eff}$. In the limit,
the generalized Schwarzschild geometry (\ref{5des-7}) becomes
\bea
ds^2&=&{{du^2}\over{u^2}}+u^2 \Bigg ( -dt^2 + \left (1+{{Q_{\rm
eff}^2}\over{2r}}\right )^6
\left ( 1-{{Q_{\rm eff}^2}\over{2r}}\right )^{-2}d\rho^2 \nonumber\\
&&\qquad\qquad\qquad + \left [ \left ( 1+{{Q_{\rm
eff}^2}\over{2r}}\right )^4 + {{2Q_{\rm eff}^2}\over{r}}\left (
1+{{Q_{\rm eff}^2}\over{2r}}\right )^{-2} \right ] \rho^2 d\Omega^2
\ \Bigg ) \ .\label{5des-10}
\eea
In the limit ${\tilde E}^2=-2Q_{\rm eff}^2/r$. Then, the extremal black hole geometry can be
approximated to yield \bea ds^2&=&{{du^2}\over{u^2}}+u^2\left [
-dt^2+\left(1-{{Q_{\rm
eff}^2}\over{2r}}\right)^{-2}d\rho^2+\rho^2d\Omega^2 \right ]
\label{5des-11} \eea It is important to note that the extremal black
hole geometry obtained following a noncommutative $D_3$-brane
\cite{km-2} is a precise generalization of that obtained in a $4D$
effective string theory\cite{garfinkle}. The effective parameter
$Q^2_{\rm eff}$ in eq.(\ref{5des-11}) may be interpretated as a
light mass of the extremal black hole. Thus in the limit,
the extremal solution
is well approximated by the $AdS_5$ geometry and is given by \be
ds^2= - u^2\ dt^2 + {{du^2}\over{u^2}} + u^2 d\Omega_3^2\ ,
\label{5des-12} \ee where $d\Omega_3^2$ governs the $S^3$ geometry.
The analysis suggests that the generalized Schwarzschild black
hole (\ref{5des-7}) in its gravity decoupling limit describes an
$AdS_5$ geometry, which is in agreement with the
$AdS/CFT$-correspondance established in string theory.

\section{Extremal black hole geometries in $4D$}

\subsection{Einstein metric}

Let us consider the Kaluza-Klein compactification of the effective
action (\ref{5des-4}) obtained in string bulk in Einstein frame. The
$5D$ metric may be given explicitly in terms of the $4D$ metric
$G_{\mu\nu}(x)$, $U(1)$ gauge field $A_{\mu}(x)$ and a scalar $\phi(x)$. It
takes a form \bea {\cal G}^E_{\a\b}=e^{{2\phi}/{\sqrt{3}}} \left
(\matrix {{G^E_{\mu\nu} + e^{-2{\sqrt{3}}\phi}A_{\mu}A_{\nu}} &
{e^{-2{\sqrt{3}}\phi}A_{\mu}} \cr {e^{-2{\sqrt{3}}\phi}A_{\nu}} &
{e^{-2{\sqrt{3}}\phi}}} \right )\ .\label{4ds-1} \eea

\par
\sp
Now, the Kaluza-Klein compactification of the $5D$ effective action
(\ref{5des-4}) is performed. Ignoring the gauge Chern-Simon terms,
the $4D$ action becomes
\be
S=\int d^4 x {\sqrt{-G^E}}\ \left [ R\ -
2(\partial\phi)^2 \ - {1\over3}(\partial\Phi)^2\ - {1\over4}
F^{(i)}D_{ij}[\Phi,\phi]{F^{(j)}}\right] \ ,\label{4ds-2}
\ee
The relevant vector field multiplet is
\bea
F^{(i)}_{\mu\nu}=\left( \matrix{{F^1_{\mu\nu}} \cr {F^{(m)}_{\mu\nu}}}
\right) \ .\label{4ds-4}
\eea
The moduli matrix is given by
\be
{D}_{ij}[\Phi,\phi]=\ \left( \matrix { {e^{-2{\sqrt{3}}\phi}} & {0}
\cr {0} & {e^{-2[2\Phi+ {\sqrt{3}}\phi]/3}\Lambda_{mn}} }\right ) \ .\label{4ds-5}
\ee
The equations of motion for the
gravity, the scalars and the gauge fields are, respectively, given by \bea
&&R_{\mu\nu}=\ 2\partial_{\mu}\phi \partial_{\mu}\phi\ +
{1\over3}\partial_{\mu}\Phi\partial_{\nu}\Phi +\
{1\over{2}}D_{ij}[\Phi,\phi]\left (
F^{(i)}_{\mu\l}{F^{(j)\l}_{\nu}}-{{1}\over{4}}F^{(i)}{F^{(j)}}
\right )\ ,\nonumber\\
&&\partial_{\mu}\left(\sqrt{-G^{E}} \
\partial^{\mu}\phi\right)={1\over{16}}{\sqrt{-G^E}}\ {{\partial
D_{ij}}[\Phi,\phi]\over{\partial\phi}}F^{(i)}F^{(j)} \ , \nonumber\\
&&\partial_{\mu}\left(\sqrt{-G^E}\
\partial^{\mu}\Phi\right)={{3}\over{8}}{\sqrt{-G^E}}\ {{\partial D_{ij}[\Phi,\phi]}\over{\partial\Phi}}F^{(i)}F^{(j)}
\nonumber\\{\rm and}
&&\quad \partial_{\mu}\left (\sqrt{G^{E}}\
{D}_{ij}[\Phi,\phi]F^{(j)\mu\nu}\right)=0 \ .\label{4ds-8}
\eea
The most general static, spherically symmetric, solutions to these equations can be given by
\be
ds^2= - a^2(r) \ dt^2 +\ a^{-2}(r) \ dr^2 + \
b^2(r)\ d\Omega^2 \ .\label{4ds-9}
\ee
Let us first consider the
case in presence of a magnetic field only. Then, the EM-fields are
\be
F^{(i)}=\ Q^{m(i)}_{\rm eff} \ \sin\theta \ d\theta\wedge d\phi \ ,\label{4ds-10}
\ee
where $Q^{m(i)}_{\rm eff}$ denote the effective
magnetic charges for $i=(1,2,3 \dots )$. The independent nonzero components of
the Ricci tensor, in an orthonormal basis, are computed to yield
\bea
&&R_{tt}=\ -{{a^2}\over{b^4}}V_{\rm eff}(\Phi,\phi)\ ,\nonumber\\
&&R_{rr}= \ 2 ( \partial_r\phi)^2\ + {1\over{a^2b^4}} V_{\rm eff}(\Phi,\phi)\nonumber\\
{\rm and}&&R_{\theta\theta}=\ -{1\over{b^2}}V_{\rm eff}(\Phi,\phi)\
,\label{4ds-11}
\eea
where $V_{\rm eff}(\Phi,\phi)$ signifies the
interaction between the moduli and gauge fields in the effective
string theory. It can be expressed as
\be
V_{\rm eff}(\Phi,\phi)=\ -
{1\over4}Q^{m(i)}{D}_{ij}[\Phi,\phi]Q^{m(j)}\ .\label{4ds-12}
\ee
For simplicity, we consider a constant $\Phi$, i.e. $V_{\rm
eff}(\Phi,\phi)|_{\Phi=const.}\rightarrow\tilde{V}_{\rm eff}(\phi)$. Then, the
eqs.(\ref{4ds-11}) are worked out using the arbitrary metric (\ref{4ds-9}).
They simplify drastically to yield
\bea
&&R_{rr} +\ {1\over{a^4}}R_{tt} =\ 2(\partial_r\phi)^2
\nonumber\\ {\rm and}
&&\partial_r^2(a^2b^2)=2 \ .\label{4ds-13}
\eea
Further simplification gives rise to the following relations:
\bea
&&(\partial_r\phi)^2\ =\ - {1\over{b}} \partial_r^2b \nonumber\\
{\rm and}&&  (a^2b^3)\ \partial^2_r b \ + \ ab^2\Big [ a +
\partial_r a\Big ]\partial_r b^2\ - b^2\ = \tilde{V}_{\rm
eff}(\phi)\ . \label{4ds-14}
\eea
The arbitrary functions $a(r)$ and $b(r)$ are worked out from the above relations to yield a charged
black hole solution in the effective string theory. A priori, it can
be written as
\be
ds^2=\ -\left(1-{{2M_{\rm
eff}}\over{r}}\right)dt^2+\left(1-{{2M_{\rm
eff}}\over{r}}\right)^{-1}dr^2\ + \ r\left( r\ -{{Q_{\rm
eff}^2e^{-2\phi_h}}\over{2M_{\rm eff}}}\right)d\Omega^2\
,\label{4ds-15}
\ee
where $\phi_h$ is a constant value of the scalar
field $\phi$ and shall be identified with its value on the event horizon. It satisfies
\be
e^{2\phi}=e^{2\phi_h}\left(1-{{Q_{\rm eff}^2 e^{-2\phi_h}}\over{2rM_{\rm eff}}}\right) \ .\label{4ds-16}
\ee
For instance when $Q=0$, the charged black hole reduces to the
Schwarzschild geometry. However for $Q\neq 0$, the radius of $S^2$
for a constant $r$ and $t$ depends on $Q_{\rm eff}$. There, the area
of $S^2$ is reduced in comparison to that of the Schwarzschild black
hole. Importantly, the above black hole solution precisely resembles
to that obtained in string theory \cite{garfinkle}. At this point,
we recall the fact that the charged black hole geometry
(\ref{4ds-15}) is obtained in presence of a noncommutative
$D_3$-brane. Since the effective string theory is defined in the
gravity decoupling limit, the charged black hole describes the near
horizon geometry. Thus, the correct black hole geometry in the
frame-work is governed by
\be ds^2=\ -\left(1-{{2Q^2_{\rm
eff}}\over{r}}\right)dt^2+\left(1-{{2Q^2_{\rm
eff}}\over{r}}\right)^{-1}dr^2\ + \ r\left( r\
-{1\over2}e^{-2\phi_h}\right ) d\Omega^2\ .\label{4ds-17}
\ee
In the case, the effective potential (\ref{4ds-12}) takes a simplified
form
\be
V_{\rm eff}=\ \left(r- {1\over2} e^{-2\phi_h}\right)^2 .
\label{4ds-18}
\ee
It implies that the area of the horizon is
reduced due to the nonzero constant value of the moduli field there.
In other words, the area of the event horizon is affected in
presence of the $V_{\rm eff}(\Phi,\phi)$ in the string frame-work.

\subsection{String frame}

In this section, we perform a similar analysis with the string metric,
in presence of both non-zero electric and magnetic fields. We
re-scale the string metric $G_{\mu\nu}=\ e^{-2\phi}G^E_{\mu\nu}$.
The $4D$ effective string action in Einstein frame becomes
\bea
S=\int d^4 x {\sqrt{-G}}\ e^{-2\phi} \left [ R + \ 4
(\partial\phi)^2 -\ {1\over3} (\partial\Phi)^2 -\ {1\over4}
F^{(i)}{\tilde D}_{ij}[\Phi,\phi]{F^{(j)}}\right ]\ .\label{4dss-1}
\eea
The moduli field in the case is
\be
{\tilde D}_{ij}[\Phi,\phi]=\ \left (\matrix{ {e^{-2[\sqrt{3}+1]\phi}} & {0} \cr {0} &
{e^{-2[(\sqrt{3}+3)\phi + 2\Phi]/3}\Lambda_{mn}}  }\right )\ .\label{4dss-3}
\ee
The equations of motion are
\bea
&&R_{\mu\nu} =\ - 4\ \partial_{\mu}\phi\partial_{\nu} \phi\ +\
{1\over3}\partial_{\mu}\Phi\partial_{\nu}\Phi\ +\ {1\over2}{\tilde
D}_{ij}[\Phi,\phi]\left
(F^{(i)}_{\mu\lambda}F^{(j)\lambda}_{\nu}-{1\over4} G_{\mu\nu}F^{(i)}F^{(j)}\right )\ ,\nonumber\\
&& \partial_{\mu}\left(\sqrt{-G}\partial^{\mu}\phi \right ) =\
{1\over{16}} {\sqrt{G}}\left [ F^{(i)}\left( {\tilde
D}_{ij}[\Phi,\phi]-{{1}\over{2}}{{\partial {\tilde D}_{ij}[\Phi,\phi]}\over{\partial\phi}}
\right){F^{(j)}}\right ]\ ,\nonumber\\
&&{\partial_{\mu}\left(\sqrt{-G}\ \partial^{\mu}\Phi\right)}
=\ {3\over8}{\sqrt{-G}}\ {{\partial {\tilde D}_{ij}[\Phi,\phi]}\over{\partial\Phi}}F^{(i)}{F^{(j)}}\nonumber\\
{\rm and}\qquad &&\partial_{\mu}\left(\sqrt{-G}e^{-2\phi} {\tilde
D}_{ij}[\Phi,\phi]{F^{(j)\mu\nu}}\right)=0\ .\label{4dss-7}
\eea
Let us consider an arbitrary metric ansatz (\ref{4ds-9}) in
presence of both electric and magnetic field in the theory.
They are given by
\be
F^{(i)}={{Q^{e(i)}_{\rm eff}}\over{b^2}}e^{2\phi} dt\wedge dr\
+\ Q^{m (i)}_{\rm eff} \sin\theta\ d\theta \wedge d\phi \
.\label{4dss-8}
\ee
The non-zero components of Ricci tensor are
worked out using eqs.(\ref{4ds-9}) and (\ref{4dss-7}). They satisfy
the following relations: \bea
&&R_{tt} = \ -{{a^2}\over{b^4}}V_{\rm eff}(\Phi,\phi)\ ,\nonumber\\
&&R_{rr}\ + 4\ (\partial_r\phi)^2= {1\over{a^2b^4}} V_{\rm
eff}(\Phi,\phi)\nonumber\\
{\rm and}\qquad &&R_{\theta\theta}=\ -{{1}\over{b^2}}V_{\rm
eff}(\Phi,\phi).\label{4dss-9} \eea where \be V_{\rm eff}(\Phi,\phi)
=\ - {1\over4}\left (Q^{m(i)}_{\rm eff}{\tilde
D}_{ij}[\Phi,\phi]Q^{m(j)}_{\rm eff}\ +\ Q^{e(i)}_{\rm eff}{\tilde
D}_{ij}[\Phi,\phi]e^{4\phi}Q^{e(j)}_{\rm eff}\right )\ .\label{4dss-10} \ee
Further simplification yields
\bea
&&\partial^2_r(a^2b^2) =\ 2 \ ,\qquad {1\over{b}}\partial_r^2b =\ 2 (\partial_r\phi)^2\nonumber\\
{\rm and}\quad &&2ab^3 (\partial_ra) (\partial_rb)\ + \ a^2b^2
(\partial_r b)^2 -\ b^2 =\ \tilde{V}_{\rm eff}(\phi) +\ 2b^2
(\partial_r\phi)^2 \ .\label{4dss-13}
\eea
Then, the black hole solution in the string-frame can be given by
\be
ds^2= -\ {{r^2}\over{{\bar Q}^2_{\rm eff}}} dt^2 +\ {{{\bar Q}^2_{\rm
eff}}\over{r^2}} dr^2 +\ {\bar Q}^2_{\rm eff} d\Omega^2 \ ,
\label{4dss-14} \ee where ${\bar Q}_{\rm eff}$ is the total
effective charge due to all the electric and magnetic fields in the
frame-work. As described in the Einstein-frame, the $4D$ solution
(\ref{4dss-14}) in the string frame describes a near horizon
geometry $AdS_2\times S^2$ of the  Reissner-Nordstrom solution.

\par
\sp
The stability analysis of the extremal black hole geometries
(\ref{4ds-17}) and (\ref{4dss-14}) are performed by taking into
account the attractor behaviour of the geometries at its event
horizon. The total charge ${\bar Q}_{\rm eff}$ is computed using the
attractor mechanism, which can be expressed in terms of the charges
associated with the gauge fields derived from the two-forms and a
three-form in $5D$.

\subsection{Black hole entropy}

Let us consider the $4D$ extremal black hole solution obtained in
the string frame (\ref{4dss-14}). Since the geometry $AdS_2\times
S^2$ is obtained in the gravity decoupling limit, it can be seen to
be associated with two different length scales in a noncommutative
farme-work \cite{kar-panda,km-1,km-2}. If $l_{\perp}$ and $l_L$ are,
respectively, the $\perp$- and $L$- length scales,
then the extremal black hole geometry can be re-expressed as
\be
ds^2 = l_{\perp}^2 \left (-\ r^2 dt^2\ + {{dr^2}\over{r^2}} \right )\ + l_L^2\
d\Omega^2 \ .\label{BH-1}
\ee
The appropriate EM-field components
may be obtained from (\ref{4dss-8}). They are \be F_{rt}^{(i)}=
E^{(i)}\qquad {\rm and}\qquad
F_{\theta\phi}^{(i)}=
{1\over{2\pi}} B^{(i)} \sin \theta \ ,\label{BH-2}
\ee
where $E^{(i)}$ and $B^{(i)}$ are the electric and magnetic fields respectively.
Since the $\perp$- and $L$- spaces in the effective geometry are
scaled apart, they give rise to two non-vanishing components of the
Riemann tensor. They are
\bea
&&R_{\a\b\gamma\delta} = {{R_{rtrt}}\over{det(G_{rt})}}(G_{\a\gamma}G_{\b\delta}-G_{\a\delta}G_{\b\gamma})\nonumber\\
{\rm and}\qquad
&&R_{mnpq}={{R_{\theta\phi\theta\phi}}\over{det(G_{\theta\phi})}}(G_{mp}G_{nq}-G_{mq}G_{np})
\ ,\label{BH-3}
\eea
where $(\a,\b,\gamma,\delta)$ specify the
$\perp$-space and $(m,n,p,q)$ there describe the $L$-space. The effective potential can
be checked to yield
\be
V_{\rm eff}=\ -{{\det(G_{rt})}\over{R_{rtrt}}}\ =
\ {{\det(G_{\theta\phi})}\over{R_{\theta\phi\theta\phi}}}\ . \label{BH-4}
\ee
Now the entropy function $f(l_{\perp},l_L, E^{(i)}, B^{(i)},
\phi)$ of the $4D$ extremal black hole (\ref{BH-1}) is given by
\cite{sen}
\be
f(l_{\perp},l_L,E^{(i)}, B^{(i)},\phi)\ = \int d\theta d\phi\ \sqrt{-G}\
{\cal{L}}\ ,\label{BH-5}
\ee
where ${\cal L}$ denotes the lagrangian
density in eq.(\ref{4dss-1}). Using the on-shell condition, the
action is simplified and can be expressed in terms of gauge fields
only. Then, the attractor mechanism for the extremal black hole
(\ref{BH-1}) at its event horizon radius $r_h$ is worked out to
obtain the entropy function. Explicitly, it takes a form
\be
f(l_{\perp},l_L,E^{(i)}, B^{(i)}, \phi)\ = {{4\pi l_L}\over{l_1}} \Big [E^{(i)}
{\tilde D}_{ij}(\phi) E^{(j)}\Big
]_{\phi\rightarrow\phi_h}\ .\label{BH-6}
\ee
The Legendre transform of the entropy function becomes
\be
 S_{BH}=2\pi \Big ( E^{(i)}{{\partial f}\over{\partial E^{(i)}}} - f\Big )
\ . \nonumber
\ee
Then the black hole entropy can be computed to yield
\bea
S_{BH}\ =
{1\over4} \Big [ 4\pi \tilde{V}_{\rm eff}(\phi)\Big
]_{\phi\rightarrow \phi_h}\ .\label{BH-7}
\eea
where $\tilde{V}_{\rm eff}(\phi)$ on the event horizon of the extremal black hole defines
$r_h^2$. It is in agreement with the Bekenstein-Hawking area law for
a black hole.

\section{Concluding remarks}

To conclude, we have revisted the generalized RN- and Schwarzschild black hole geometries in $4D$, recently obtained by us \cite{km-2}, following a noncommutative $D_3$-brane frame-work.
Apart from the fact that the generalized RN-solution in string theory is new, it has provided
a forum to investigate some aspects of quantum gravity. In the gravity decoupling limit, these black holes coincide to yield a Schwarzschild geometry and its mass term is shown to be associated with the $1/r^4$. This in turn prompts one to believe for the existence
of an $7D$ effective theory in the decoupling regime. The noncommutative scaling in the frame-work is exploited to conlcude that the effective space-time turns out to be $5D$ instead of $7D$.

\sp
\par
In the context, a relevant effective  theory in $5D$ was obtained in type IIB string theory.
The coupling of moduli fields to the gauge field strengths has been incorporated in terms of an effective potential in the theory. The black hole solutions in $5D$ were worked out with a
static, spherically symmetric metric ansatz in presence of an arbitrary electric field ${\tilde E}$. These black holes were argued to describe the extremal geometries
$AdS_2\times S_3$ in the gravity decoupling limit. It was shown that a black hole solution is
in precise agreement with the gravity dual of a $D_3$-brane in presence of a constant magnetic field \cite{li-wu}. As a result, our result provides evidence for a holographic correspondence
between the boundary noncommutative gauge theory and the bulk of the string. The extremal limit
was further analyzed to conlude an $AdS_5$ geometry in the bulk. Interestingly, the near horizon geometry of a generalized RN-black hole in $5D$ was shown to be a higher dimensional
generalization of the charged black hole obtained in effective string theory \cite{garfinkle}
In fact, our analysis provides evidences to the strong-week copupling duality between the noncommutative gauge and the ordinary gravity sector in the theory

\sp
\par
In order to compare the black hole geometries in a noncommutative $D_3$-brane to that in effective string theory, Kaluza-Klein compactification was performed in $5D$. The relevant black hole geometries were obtained in Einstein and string frames in presence of an effective
potential $V_{\rm eff}$. The potential was shown to be characterized by the EM-charges and
is independent of the value of moduli field there. Attractor mechanism was adopted to compute
the entropy function of the black hole in a noncommutative frame-work. Then, the entropy of a macroscopic black hole was expressed in terms of its efffective potential at the event horizon

\sp
\par
It is important to keep a note that, in the gravity decoupling limit the generalized black hole reduces to an appropariate $AdS$ geometry in the noncommutative frame-work. As the limit is intrinsic to the frame-work, it plays a vital role. For instance, the limit may be interpreted as the one leading to $AdS$ boundary in the theory. Now, let us recall the behaviour of Hawking temperature \cite{km-2}, $i.e.$ it decreases to zero with an increase in Hawking radiation for a GRN-black hole and finally increases to atatin the Hagedorn temperature for a Schwarzschild geometry  Taking into account the variation of Hawking temperature, the event horizon can be
seen to be stretched between the GRN- and Schwarzschild geometries in the frame-work. Since a
noncommutative $D_3$-brane is known to govern the event horizon of a black hole, the stretch at
the eveny horizon can be interpreted as due to the noncommutative or new geometry there.
Since the $D_3$-brane can be described by an appropriate $AdS$ geometry, it would be interesting to check the noncommutative formalism in presence of a cosmological constant.

\sp
\sp
\sp

\noindent
{\large\bf Acknowledgements:}

\sp
S.K. thanks the high energy section at the Abdus Salam I.C.T.P, Trieste for its support,
where a part of this work is performed.
The work of S.K. is partly supported, under a SERC fast track young scientist PSA-09/2002
research project, from D.S.T, Govt.of India. The work of S.M. is partly supported by a
C.S.I.R. research fellowship, New Delhi.

\sp
\sp
\sp

\def\anp{Ann. of Phys.}
\def\cmp{Commun. Math. Phys.}
\def\prl{Phys. Rev. Lett.}
\def\prd#1{{Phys. Rev.} {\bf D#1}}
\def\jhep{J.High Energy Phys.}
\def\cqg{{Class. \& Quantum Grav.}}
\def\plb#1{{Phys. Lett.} {\bf B#1}}
\def\npb#1{{Nucl. Phys.} {\bf B#1}}
\def\mpl#1{{Mod. Phys. Lett} {\bf A#1}}
\def\ijmpa#1{{Int. J. Mod. Phys.} {\bf A#1}}
\def\ijmpd#1{{Int. J. Mod. Phys.} {\bf D#1}}
\def\rmp#1{{Rev. Mod. Phys.} {\bf 68#1}}


\end{document}